\documentclass{iau}
\usepackage{graphicx}

\newcommand{\grays}{$\gamma$-rays}
\newcommand{\gray}{$\gamma$-ray}
\newcommand{\gsim}{\mbox{\hspace{.2em}\raisebox{.5ex}{$>$}\hspace{-.8em}\raisebox{-.5ex}{$\sim$}\hspace{.2em}}}

\newcommand{\um}{$\mu$m}
\newcommand{\cm}{\,{\rm cm}}    \newcommand{\km}{\,{\rm km}}
\newcommand{\ps}{\,{\rm s}^{-1}}
\newcommand{\twCO}{$^{12}$CO}   \newcommand{\thCO}{$^{13}$CO}
\newcommand{\Jotz}{$J$=1--0}    \newcommand{\Jtto}{$J$=2--1}
\newcommand{\VLSR}{V_{\rm LSR}} \newcommand{\Msun}{M_{\odot}}
\newcommand{\Rb}{R_{\rm b}}     \newcommand{\pu}{p_5}

\title[Molecular Environments of SNRs] 
{Molecular Environments of Supernova Remnants
\thanks{Supported by the 973 Program grant 2009CB824800,
   the NSFC grants 11233001, 11103082, and 11203013,
   grants 20120091110048 and 20110091120001
   from the Educational Ministry of China,
   and grant 2011M500963 from the China Postdoctoral Science Foundation.}
}

\author[Chen, Y.\ et al.]   
{Yang Chen$^{1,2}$, Bing Jiang$^1$, Ping Zhou$^1$, Yang Su$^3$, Xin Zhou$^{3,2}$,\\
 Hui Li$^{4,1}$, \and Xiao Zhang$^1$
}

\affiliation{
$^1$ Department of Astronomy, Nanjing University, Nanjing 210093, China \\
 email: {\tt ygchen@nju.edu.cn,
 bjiang@nju.edu.cn, pingzhou@nju.edu.cn, zxmysky@163.com}\\[\affilskip]
$^2$ Key Laboratory of Modern Astronomy and Astrophysics,
 Nanjing University, Ministry of Education, Nanjing 210093, China\\[\affilskip]
$^3$ Purple Mountain Observatory, 2 West Beijing Road, Nanjing 210008, China\\
 email: {\tt yangsu@pmo.ac.cn, xinzhou@pmo.ac.cn} \\[\affilskip]
$^4$ Department of Astronomy, University of Michigan, 500 Church Street, Ann Arbor, MI~48109, USA\\
 email: {\tt hliastro@umich.edu}
}

\pubyear{2013}
\volume{xxx}  
\pagerange{??--??}
\jname{IAU Symposium 296~~Supernova Environmental Impacts}
\editors{Alak Ray \& Dick McCray, eds.}
\begin{document}

\maketitle

\begin{abstract}
There are about 70 Galactic supernova remnants (SNRs) that are now confirmed
or suggested to be in physical contact with molecular clouds (MCs)
with six kinds of evidence of multiwavelength observations.
Recent detailed CO-line spectroscopic mappings of a series of SNRs
reveal them to be in cavities of molecular gas, implying the roles the
progenitors may have played.
We predict a linear correlation between the wind bubble sizes
of main-sequence OB stars in a molecular environment and the stellar masses and
discuss its implication for supernova progenitors. The molecular environments of SNRs
can serve as a good probe for the \grays\ arising from the hadronic interaction
of the accelerated protons, and this paper also discusses the \gray\ emission
from MCs illuminated by diffusive protons that escape from SNR shocks.

\end{abstract}

\firstsection 
\section{Introduction}

The interplay between supernova remnants (SNRs) and their molecular environments
plays an important role in many aspects of astrophysical studies.
Molecular gas takes up about half mass of Galactic interstellar medium,
and most core-collapse supernovae are believed to be located close to
giant molecular clouds (MCs) where their progenitor stars are given birth to
(e.g., \cite[Huang \& Thaddeus 1986]{HT86}).
The SNR shock waves propagating in molecular environments can compress, heat,
excite, ionize, and even dissociate molecules. They are suggested to be one of
the mechanisms of triggering star formation.
They influence the chemical evolution
of the gas and produce otherwise impossible or seldom detected molecular emission
(e.g., 1720MHz OH maser, HCO$^+$, etc.).
Shock interaction with molecular clouds can generate \grays\ as a result of
neutral pion decay after p-p collision (hadronic interaction),
which may serve as a probe of SNR shock acceleration of relativistic protons.
Once an SNR-MC association is established, the kinematic distance of the
SNR can be determined with the local standard of rest (LSR) velocity
of the MC.

This article is composed of contents on SNR-MC associations, some of our
CO-line observations, bubbles in molecular environments, and probe
for hadronic interaction.

\section{Census of SNR-MC associations}

How many among the $\gsim300$ Galactic SNRs are in physical contact with MCs?
There have been some endeavors in surveying and cataloging the SNR-MC associations.
In an early CO-line survey toward 26 outer ($l=70^{\circ}$--$210^{\circ}$)
Galactic SNRs, about half were found
spatially coincident with large MC complexes
(\cite[Huang \& Thaddeus 1986]{HT86}).
After a survey of 1720\,MHz OH maser emission towards Galactic SNRs,
\cite[Frail et al.\ (1996)]{Frail_etal1996} and
\cite[Green et al.\ (1997)]{Green_etal1997}
revealed the unique diagnostic relation between such masers and the SNR shocks
in MCs and identified nearly a score of SNRs interacting with MCs.
\cite[Seta et al.\ (1998)]{Seta_etal1998} listed 26 SNRs detected
in CO- and H$_2$-line emissions along the line of sight,
but no physical evidence of shock-MC interaction was given for them.

Recently, we presented a catalog of the interacting SNRs,
with a summary of six kinds of multiwavelength observational
evidence for judging the contact of SNRs with MCs
(\cite[Jiang et al.\ 2010]{Jiang_etal2010}).
Among the six observational evidences, firstly, the 1720\,MHz OH
maser is the most important, which is presently widely
accepted as a signpost of SNR shock-MC interaction.
When the molecular gas is compressed by a C-type shock to a
density of order $\sim10^5\cm^{-3}$ and
the temperature reaches 50--125~K,
collisional pumping causes the population inversion of the
hyperfine rotational level $^2\Pi_{3/2}(J=5/2)$ of OH molecules,
from which the 1720\,MHz maser arises
(\cite[Lockett et al.\ 1999]{Lockett_etal1999}).
Since this OH satellite line was noted by
\cite[Goss \& Robinson (1968)]{GR1968}, explorations and surveys
in the line have been repeatedly made toward Galactic SNRs,
and the masers have been detected from 25 SNRs.
Secondarily, molecular (CO, HCO$^+$, CS, etc.) line broadening
or asymmetric profile is another important kinematic evidence
of shock perturbation of the molecular gas
(e.g., \cite[DeNoyer 1979]{Denoyer1979}).
Line broadening is present in 18 SNRs listed in our table.
The third kind of evidence is high high-to-low excitation line
ratio in line wings (\cite[Seta et al.\ 1998]{Seta_etal1998}).
In the broad wings of the shocked \twCO, where both the
\Jtto\ and \Jotz\ emissions are optically thin, their ratio
can be $\gsim1$.
The fourth is the detection of near-infrared (IR) emission,
e.g., [Fe II] line
or ro-vibrational lines of H$_2$ (such as H$_2$ 1--0 S(1) line
at 2.12~\um\ and H$_2$ 0--0 S(0)--S(7) lines), due to shock excitation.
The fifth is the specific IR colors suggesting molecular shocks,
e.g., 3.6 \um/5.8 \um\ and 4.5 \um/8 \um\
in the Spitzer IRAC observation (see
\cite[Reach et al.\ 2006]{Reach_etal2006} and Fig.2 therein).
The sixth is the morphological agreement or correspondence of
molecular features with SNR features (e.g., arc, shell, interface,
etc.).
Besides the OH masers, a combination of the sixth (spatial) evidence
with one of the other four kinematic/physical evidences is also
regarded as convincing criteria for judging the SNR-MC interaction.
Thus, in our catalog, there are 34 SNRs ``confirmed", 11 ``probable",
and 19 ``possible" to be in physical contact with MCs.
A most recent CO survey for $l=60^\circ$--$190^\circ$ showed
additional six SNRs with their radio morphology in a good spatial
relation with MCs, without direct evidence for the interaction
(\cite[Jeong et al.\ 2012]{Jeong2012}),
thus raising the number of ``possible" interacting SNRs to 25.

\section{Our CO observations of a series of SNRs} \label{sec:ours}

The OH satellite line masers have proven to be a powerful tool for
the identification of interacting SNRs, nonetheless they cannot be used
to investigate the detailed information of molecular environments
of the SNRs.
Also, a number of SNR-MC associations may be elusive in the search
in the maser line because of the weak emission below the detection thresholds.
For this purpose, CO, other than H$_2$
(without a permanent electric dipole moment), is a practical tracer of MCs
and commonly used for case studies of SNR-MC interaction.
Here we review our recent CO observations of a series of interacting SNRs
made with the millimeter and sub-millimeter telescopes in the Purple Mountain
Observatory at Delingha, Seoul Radio Astronomy Observatory,
Koelner Observatory for Submillimeter Astronomy, and other observatories.

\subsection{Kes\,69}
Kes\,69 is morphologically characterized by the bright
radio, IR, and X-ray emissions only at the southeastern boundary.
The 1720\,MHz OH masers are detected in both the northeast and the southeast,
but with different LSR velocities
(\cite[Green et al.\ 1997]{Green_etal1997};
\cite[Hewitt et al.\ 2008]{Hewitt_etal2008}).
Our \twCO\ (\Jotz) observation (\cite[Zhou et al.\ 2009]{Zhou_etal2009})
discovers a molecular arc in the LSR velocity interval
77--$86\km\ps$ (consistent with $85\km\ps$ for the southeastern maser).
The arc is in good morphological agreement with
the multiwavelength partial shell of the SNR.
The HCO$^+$ emission is detected at the position of the radio brightness peak
also at the LSR velocity $\VLSR=85\km\ps$, which is consistent with the presence
of the southeastern maser, both resulting from C-shock interaction.
These evidences strongly suggest that Kes~69 is physically associated with
the MC at the systemic velocity $85\km\ps$.
From this velocity, a kinematic distance 5.2~kpc is derived.
We ascribe the multiwavelength emissions arising from the
southeastern partial shell of the SNR to the impact of the SNR shock
on a dense, clumpy patch of molecular gas, which is likely to be the cooled debris
of the material swept up by the progenitor's stellar wind.

\subsection{Kes\,75}
The young composite SNR Kes\,75, with a pulsar wind nebula inside,
displays only a half shell in the south.
We find that the \twCO\ (\Jotz) line profile of the $\VLSR=45$--$58\km\ps$
MC is broadened in the blue wing and a molecular shell
unveiled in this wing (45--$51\km\ps$) encloses a cavity
(\cite[Su et al.\ 2009]{Su_etal2009}).
The southern part of the molecular shell is in good morphological
agreement with the SNR's half shell shown in X-rays, mid-IR, and
radio continuum.
These spatial and kinematic evidences indicate that Kes\,75 is
physically associated with the MC at the systemic velocity $54\km\ps$.
The associated large cloud has a mass $\gsim10^4M_{\odot}$.
The distance to this SNR is thus determined to be 10.6~kpc.
The presence of the dense molecular shell is interpreted to be
due to the same reason as that of the molecular arc in Kes\,69.

\subsection{Kes\,78}
Kes\,78 is a shell-type SNR, but only the eastern half is radio bright.
There is a $86\km\ps$ OH maser at the eastern boundary
(\cite[Koralesky et al.\ 1998]{Koralesky_etal1998}).
The unidentified TeV source J1852$-$000 to the east of the remnant
is detected by the HESS \gray\
telescope\footnote{http://www.mpi-hd.mpg.de/hfm/HESS/pages/home/som/2011/02/},
which implies that this SNR may have a relation with the
very high energy emission.
Our CO line observations toward this SNR find that it is interacting
with the MC complex at the systemic velocity $\VLSR=81\km\ps$
(\cite[Zhou \& Chen 2011]{ZC2011}).
The \thCO\ emission shows the presence of a dense cloud to the east,
from which the maser may arise and which also seems crudely correspondent
to the location of the extended TeV source.
The strong \twCO\ emission in general spatially corresponds to the
eastern bright radio shell, and the spatial extent of the SNR is
consistent with a \twCO\ cavity.
\twCO\ lines are found to be broadened in some boundary regions including
the eastern maser region, and \twCO\ \Jtto/\Jotz\ ratio is generally
elevated along the SNR boundary.
These are all the kinematic signatures of shock perturbation in the
molecular gas.
The distance to Kes\,78 is estimated to be 4.8~kpc based on the
systemic LSR velocity.
An {\sl XMM}-Newton X-ray spectral analysis for the northeastern
boundary infers an age 6~kyr for the remnant.

\subsection{3C\,396}

Analogous to Kes\,75, 3C\,396 is another composite SNR containing a
pulsar wind nebula and is semi-circular-shaped, too.
We find that the western edge, bright in radio, IR, and X-rays,
is perfectly confined by a molecular wall
(of mass $\sim10^4M_{\odot}$) revealed in \twCO\ (\Jotz)
line at $\VLSR\sim84\km\ps$ (\cite[Su et al.\ 2011]{Su_etal2011}).
The CO emission gradually gets faint from west to east, which is
indicative that the eastern region is of low gas density.
Noticeably, a finger/pillar-like molecular structure
(of mass $\sim4\times10^3M_{\odot}$) in the southwest
intrudes into SNR edge. The pillar, which may have been shocked at
the tip, should be the reason why the X-ray and radio emissions get
brightened at the southwestern boundary and some IR filaments
are present there.
The evidences for the SNR-MC interaction also include the relatively
elevated \twCO\ \Jtto/\Jotz\ line ratios in the southwestern ¡°pillar tip¡±
and the molecular patch at the northwestern boundary, as well as
the redshifted \twCO\ (\Jotz\ and \Jtto) wings (86--$90\km\ps$)
of an eastern $\VLSR\sim81\km\ps$ molecular patch.
The X-ray analysis of the hot gas infers an age $\sim$3~kyr
for the remnant and the derived relative abundances of Si, S, and Ca
of the ejecta are consistent with a B1--B2 progenitor star.

\subsection{3C\,397}
3C\,397 is a thermal composite (or mixed-morphology) SNR with an abnormal
rectangular shape. Our \twCO\ (\Jotz) observation shows that the remnant
is confined in a molecular gas cavity and embedded at the edge of a
giant MC at the systemic velocity $\VLSR=32\km\ps$
(\cite[Jiang et al.\ 2010]{Jiang_etal2010}).
The column density of the environmental molecular gas has a gradient
that increases from the southeast to the northwest and is perpendicular to the
Galactic plane, in agreement with the elongation direction of the remnant.
Solid evidence for the SNR-MC interaction is provided by the
\twCO\ line broadening of the $\sim32\km\ps$ component that is detected
at the westmost boundary of the remnant, which is
consistent with the impact of the Fe rich ejecta in the region
(\cite[Jiang \& Chen 2010]{JC2010}).
The systemic velocity of the molecular gas places 3C\,397 at a
kinematic distance of 10.3~kpc.
The mean ambient molecular density is $\sim$10--30$\cm^{-3}$,
which can explain the high volume emission measure of the X-ray emitting gas.
For the westmost line-broadened region, the density of the disturbed
molecular gas is deduced to be of order $10^4\cm^{-3}$ based on the
the pressure balance between the cloud shock and the X-ray
emitting hot gas and can be ascribed to very dense clumps,
implying a multi-phase gas environment there.

\subsection{Kes\,79 and W\,49B}
SNRs Kes\,79 and W\,49B have also been observed and preliminarily show
interesting molecular environments.

\begin{figure}[t]
\begin{center}
 \includegraphics[height=2.0in]{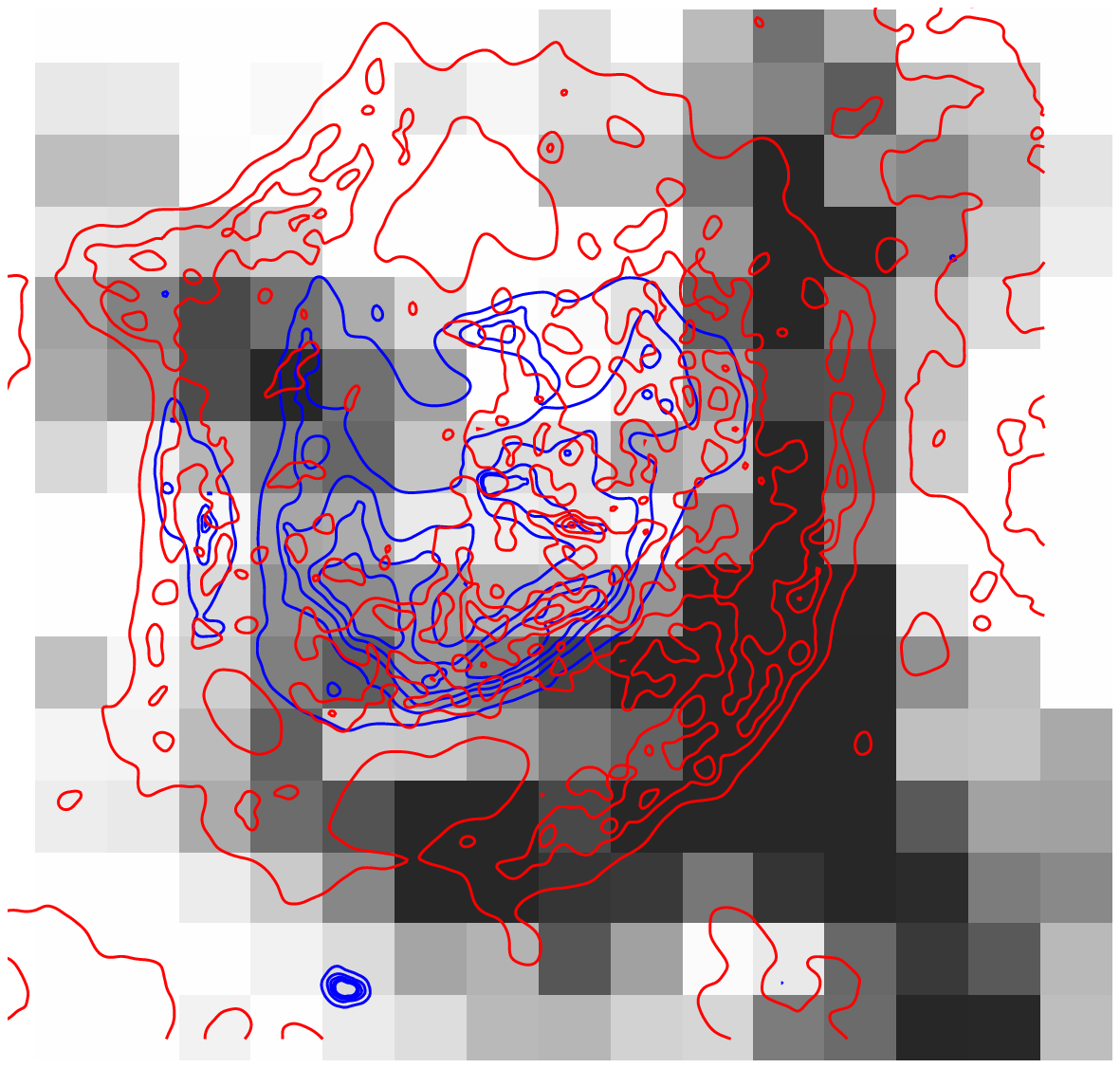}\hspace{1cm}
 \includegraphics[height=2.0in]{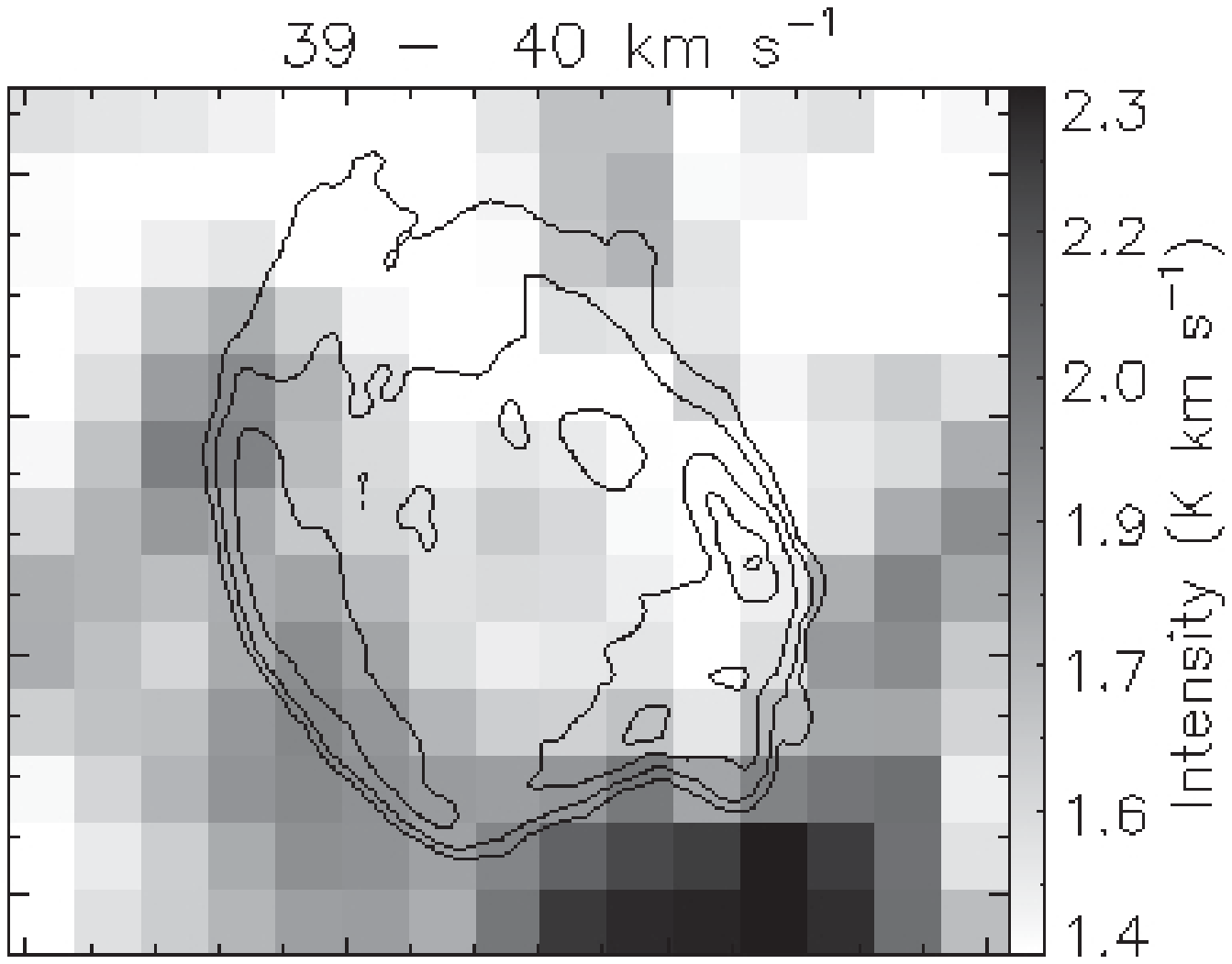}
 \caption{
 Left: \twCO\ (\Jtto) emission map at $\VLSR=112$--$115\km\ps$ for
 Kes\,79, overlaid with radio continuum
 contours (red) and X-ray contours (blue).
 The size of each grid cell is 1~arcmin.
 Right: \twCO\ (\Jtto) emission map at $\VLSR=39$--$40\km\ps$ for
 W\,49B, overlaid with radio continuum contours.
 The size of each grid cell is 0.5~arcmin.
  }
   \label{fig1}
\end{center}
\end{figure}

Kes\,79, harboring a central compact object, was demonstrated by
the {\sl Chandra} observation to have
multi-shell structure, which implies a complicated surrounding
environment (\cite[Sun et al.\ 2004]{Sun_etal2004}).
Our CO observation finds that this remnant is spatially coincident
with a molecular cavity at $\VLSR\sim113\km\ps$
(Fig.\,\ref{fig1}).
The \twCO\ \Jtto/\Jotz\ ratio is also found to be elevated,
as high as $\gsim1$, in some boundary regions.

W\,49B, sorted as a thermal composite, has an incomplete radio
shell with centrally brightened thermal X-rays.
A barrel-shaped structure with coaxial rings was revealed in the
1.64~\um\ [Fe II] image, and a strip of 2.12~\um\ shocked
H$_2$ emission extends outside of the [Fe II] emission to the southeast;
the X-ray jet-like structure is along the axis of the barrel
(\cite[Keohane et al.\ 2007]{Keohane_etal2007}).
It has been suggested that W\,49B originated inside a wind-blown bubble
interior to a dense MC.
By CO observation, we have noticed that this SNR seems to be
coincident with a molecular cavity at $\VLSR\sim39\km\ps$
(Fig.\,\ref{fig1}),
but no robust kinematic evidence has been obtained yet.
If this SNR-MC association is true, the kinematic distance to W\,49B
is 9.3~kpc.

The above observations show a common phenomenon that these interacting
SNRs all evolve in molecular cavities/bubbles, which may be created by their
progenitors' activities (i.e., most probably, stellar winds).
The cavities/bubbles are very unlikely to be produced by the SNR shocks,
because the supernova explosion energy would otherwise be much higher than
the canonical budget $10^{51}$~erg.

\section{Bubbles in molecular environment}

A bubble blown by a main sequence star achieves the maximum size
when the bubble is in pressure equilibrium with the ambient medium
(\cite[Chevalier 1999]{Ch99}).
\cite[Chen et al.~(2013)]{Chen_etal2013} show a
linear relation
for the bubble size $\Rb$ for main-sequence OB stars:
$ p^{1/3}\Rb \propto M$,
where $p$ is the pressure of the surrounding interclump
medium and $M$ the stellar mass.
Actually, by inserting parameters that are observationally determined
and model-estimated into \cite[Chevalier's (1999)]{Ch99} formula, they
find that, for 15 exemplified main-sequence OB stars, $p^{1/3}\Rb$
does correlate linearly with $M$,
and a good regression for the relation can be obtained as
\begin{equation} \label{eq:regression}
\pu^{1/3}\Rb=\alpha(M/\Msun)-\beta
 \hspace{2mm}\mbox{pc},
\end{equation}
where $\pu \equiv (p/k)/(10^5\cm^{-3}\,\mbox{K})$,
 $\alpha=1.22\pm0.05$, and $\beta=9.16\pm1.77$.

In a giant MC, the mean pressure is $\pu\sim1$
(\cite[Krumholz et al.\ 2009]{Krumholz_etal2009}), and
this number is indeed needed to confine the dense clumps and supports
the cloud against gravitational collapse
(\cite[Blitz 1993]{Bl93}; \cite[Chevalier 1999]{Ch99}).

Eq.~(\ref{eq:regression}) provides a powerful way to assess the
initial masses of the progenitors for SNRs associated with molecular
clouds.
For an initial mass below 25--$30\Msun$, the stars will terminate their lives
in Type~II-P supernova (SN) explosion, which takes up the greatest majority
of SNe~II,
after the red supergiant (RSG) phase,
without further launching Wolf-Rayet winds.
The post-main sequence stellar winds reach an extent much smaller than
the main-sequence bubbles.
The SN shocks will rapidly pass through the circumstellar
material and impact the massive shells of the bubbles,
with drastic deceleration (\cite[Chen et al.~2003]{CZWW03}).
Thus, such SNRs reflect mostly the bubble sizes, and the sizes can be
used to infer the progenitors' masses.

Table~\ref{tab1} lists the interacting SNRs that are known or suggested to
have molecular shells or be in molecular cavities, which include
the Vela SNR and RX\,J1713$-$3946
in addition to those described in \S\ref{sec:ours}.
The Vela SNR coincides with a molecular void at a velocity range of
$\VLSR$ =  $-5$ to $85\km\ps$ (\cite[Moriguchi et al.\ 2001]{Metal01}).
It is suggested that the molecular clumps are
pre-existent, rather than having been swept up by the SNR shock, and
that the SNR may have been expanding in a low density medium.
RX~J1713.7$-$3946 appears to be confined in a molecular gas cavity at
$\VLSR\sim-11$ to $-3\km\ps$
(\cite[Fukui et al.\ 2003]{Fetal03}; \cite[Moriguchi et al.\ 2005]{Metal05}).
It is suggested that this SNR is still in the free expansion phase
and the non-decelerated blast wave is colliding with the dense molecular
gas after it traveled in a low-density cavity that perhaps was produced
by the stellar wind or pre-existing supernovae
(\cite[Fukui et al.\ 2003]{Fetal03}).

The estimates of progenitors' masses for the SNRs are given in the last
column,
on the assumption that the interclump pressure to be a constant: $\pu\approx1$.
Some of the estimates can be compared with other available independent
assessments, which show considerably good consistency.
For 3C\,396, \cite[Su et al.\ (2011)]{Su_etal2011}, based on {\sl Chandra}
observation, derived a progenitor mass of 13--$15\Msun$ from
the metal abundances of the X-ray-emitting
gas that may be dominated by the SN ejecta.
For 3C\,397, a recent {\sl XMM}-Newton X-ray study has analyzed the
metal abundances of the SN ejecta and thus assessed the progenitor's mass
to be $11$--$15\Msun$ (Safi-Harb et al.\ in preparation).
For the Vela SNR, an SN~II-P progenitor mass was suggested to be
15--$20\Msun$ based on the moderate size of the wind-blown bubble
(\cite[Gvaramadze 1999]{Gv99}).

\begin{table}
  \begin{center}
  \caption{Galactic SNRs with molecular shells/in molecular cavities}  \label{tab1}
 {\scriptsize
  \begin{tabular}{|c|c|c|c|}\hline
{\bf SNR$^{\rm ref.}$} &{\bf Distance [kpc]} & {\bf $\Rb$ [pc]} & {\bf $M$ [$\Msun$]} \\ \hline
G21.8$-$0.6 (Kes\,69)$^1$ & 6.2 & 13 & $\sim18$ \\ \hline
G29.7$-$0.3 (Kes\,75)$^2$ & 10.6 & 6 &  $\sim12$ \\ \hline
G32.8$-$0.1 (Kes\,78)$^3$ & 4.8 & 17 & $\sim21$ \\ \hline
G33.6$+$0.1 (Kes\,79)$^4$ & 7 & 8 & $\sim14$ \\ \hline
G39.2$-$0.3 (3C\,396)$^5$ & 6.2 & 7 & $\sim13$ \\ \hline
G41.1$-$0.3 (3C\,397)$^6$ & 10.3 & 4.5--7 & $\sim12$ \\ \hline
G43.3$-$0.2 (W\,49B)$^7$ (?)  & 9.3 & 7 & $\sim13$ \\ \hline
G54.4$-$0.3 (HC\,40)$^8$ & 3/7 & 18/43 & $\sim22/?$ \\ \hline
G263.9$-$3.3 (Vela)$^{9, 19}$ (?) & 0.29 & 14--19 & $\sim21$ \\ \hline
G347.3$-$0.5 (RX\,J1713$-$3946)$^{11,12}$ & 1.1 & 9 & $\sim15$ \\ \hline
  \end{tabular}
  }
 \end{center}
\vspace{1mm}
 \scriptsize{
 {\it Notes:} The ``?" symbol means that the SNR-MC association needs
 to be confirmed. \\
  References---1: \cite[Zhou et al.\ (2009)]{Zhou_etal2009};
2: \cite[Su et al.\ (2009)]{Su_etal2009};
3: \cite[Zhou \& Chen (2011)]{ZC2011};
4: \cite[Sun et al.\ (2004)]{Sun_etal2004}
5: \cite[Su et al.\ (2011)]{Su_etal2011};
6: \cite[Jiang \& Chen(2010)]{JC2010};
7: this paper
8: \cite[Junkes et al.\ (1992)]{Junkes_etal1992};
9: \cite[Moriguchi et al.\ (2001)]{Metal01};
10: \cite[Dodson et al.\ (2003)]{Detal03};
11: \cite[Fukui et al.\ (2003)]{Fetal03};
12: \cite[Moriguchi et al.\ (2005)]{Metal05} }
\end{table}

\section{Probe for hadronic interaction}
The origin of cosmic rays (CRs) has drawn more and more attention motivated
by the recent decades' X- and \gray\ observations.
It is generally believed that the CRs below the ``knee" ($3\times10^{15}$\,eV)
are of Galactic origin. The candidate sites where the Galactic CRs
are accelerated include SNRs, binaries, superbubbles, Galactic center, etc.,
and SNRs are usually regarded as the most important. Relativistic electrons
accelerated up to an energy $10^{13}$--$10^{14}$\,eV by the SNR shocks
are evidenced by the X-ray synchrotron emission arising from them.
Yet, conclusive evidence for the acceleration of relativistic protons
remains poor. An indirect way is to detect the \gray\ emission generated by
the hadronic interaction of the accelerated protons, namely, neutral pion
decay after the collision of the protons with the baryons of the environmental
dense gas. Thus, the MCs with which SNRs are associated or interacting
are a perfect probe for the hadronic process.

In the second {\sl Fermi}-LAT GeV source catalog, there are 89 sources whose
95\% confidence error ranges overlap with SNRs, and among which 45\% are
of chance coincidences but six correspond to firmly identified \gray\ emitting
SNRs (\cite[Nolan et al.\ 2012]{Nolan_etal2012}).
Among the TeV sources, about ten correspond to SNRs
(\cite[Holder 2012]{Holder2012}).
Also, a correlation has been proposed between a class of GeV-TeV \gray\ sources
that are coincident with interacting SNRs and the 1720\,MHz OH masers
(\cite[Hewitt et al.\ 2009]{Hewitt_etal2009}).
However, it is not easy to differentiate whether the \grays\ arising from
the SNRs are hadronic or leptonic emission.
Recent years, a progress has been made by analysing the ``illuminating" effect
of the nearby MCs by the SNR shock-accelerated protons
(e.g., \cite[Aharonian \& Atoyan 1996]{AA1996};
       \cite[Gabici et al.\ 2009]{Gabici_etal2009};
       \cite[Li \& Chen 2010]{LC2010}, \cite[2012]{LC2012};
       \cite[Ohira et al.\ 2011]{Ohira_etal2011}).

Among these work, \cite[Li \& Chen (2010]{LC2010}, \cite[2012)]{LC2012}
established an ``accumulative diffusion" model for
the escaping protons and naturally interpreted the GeV-TeV spectra of nine
interacting SNRs.
W28 is one of the prototype interacting SNRs, in contact with a large MC
in the north and harboring a large amount of 1720\,MHz masers
(\cite[Frail et al.\ 1994]{Frail_etal1994}).
Four TeV sources in the area are detected by the H.E.S.S.\ \gray\ telescope,
positionally coincident with the northern large MC and three small MCs
in the south (\cite[Aharonian et al.\ 2008]{Aharonian_etal2008}).
The three southern clouds seem to be outside the reach of the W28 blast wave,
and {\sl Fermi}-LAT detected no significant GeV emission for two of them
(\cite[Abdo et al.\ 2010]{Abdo_etal2010}).
They show that the various relative GeV-TeV brightness of the four \gray\
sources results from the hadronic process of the accelerated protons that
escape from the shock front and bombard the nearby MCs at different radii.
The ``illuminating" protons are considered to be an accumulation
of the diffusive protons escaping from the shock front throughout
the history of the SNR expansion. For the various distances of the
MCs from the SNR center, the resulting proton spectra can have
prominently different shapes. Adopting different centric distances,
the \gray\ spectral fit for the four sources well explains their
GeV-TeV spectral properties. It is also implied that the spectral index
2.7 at the high energy side is caused by diffusion other than
directly by acceleration, and the spectral break (from 2.1 to 2.7)
at around 1\,GeV can naturally appear due to the accumulative diffusion
effect.

This model, with improvement by incorporating the finite volume of
molecular clouds, is further applied to nine \gray\ emitting interacting
SNRs (W\,28, W\,41, W\,44, W\,49B, W\,51C, Cygnus Loop, IC\,443,
CTB\,37A and G349.7+0.2).
Like that aforementioned, the $\sim$\,GeV spectral
breaks are commonly present in these SNRs; and most of them
have a spectral ($E^2 dF/dE$) ``platform" extending from the break
to lower energies.
This refined model perfectly explains the $\sim$ GeV spectral breaks
and the ``platforms", together with the available TeV data.
It is also found that the index of diffusion coefficient $\delta$
derived from the spectral fit is in the range 0.5--0.7, analogous to
the Galactic average, and the diffusion coefficient for the CRs
around the SNRs ($\chi\sim10^{-2}$) is essentially two orders of
magnitude lower than the Galactic average, which is indicative
of the suppression of CR diffusion near SNRs.

\grays\ from molecular clouds illuminated by diffusive protons that
escape from
interacting SNRs strongly support the scenario that hadrons accelerated
by SNR shocks contribute to the Galactic CRs.
A discovery of the spectral rising below $\sim200$\,MeV for IC\,443
and W44 as an unique pion-decay signature is newly released after this
Symposium (\cite[Ackermann et al.\ 2013]{Ackermann_etal2013}).

So far, nearly a dozen of SNRs are proposed to emit hadronic
\grays, including the historical SNRs Cas~A and Tycho. Against
the previous suggestion for tenuous medium surrounding Tycho
responsible for the hadronic emission, it is argued that the \grays\ are
produced from dense ambient matter, most probably MCs, too
(see \cite[Zhang et al.\ 2013]{Zhang_etal2013}).



\end{document}